\documentclass[preprint,aps,amsmath,showpacs,showkeys,nofootinbib, 
superscriptaddress]{revtex4} 
\begin{document} 

\title{\textbf{Maximal isospin few-body systems of nucleons and $\Xi$ hyperons}}
\author{H.~Garcilazo} 
\email{humberto@esfm.ipn.mx} 
\affiliation{Escuela Superior de F\' \i sica y Matem\'aticas, \\ 
Instituto Polit\'ecnico Nacional, Edificio 9, 
07738 M\'exico D.F., Mexico} 
\author{A.~Valcarce} 
\email{valcarce@usal.es} 
\affiliation{Departamento de F\'\i sica Fundamental and IUFFyM,\\ 
Universidad de Salamanca, E-37008 Salamanca, Spain}
\author{J.~Vijande}
\email{javier.vijande@uv.es}
\affiliation{Departamento de F\'{\i}sica At\'{o}mica, Molecular y Nuclear, Universidad de Valencia (UV)
and IFIC (UV-CSIC), E-46100 Valencia, Spain}
\date{\today} 
\begin{abstract}
By using local central Yukawa-type interactions that reproduce the low-energy 
parameters of the latest updates of the Nijmegen ESC08c potentials we show 
that the $N\Xi$, $NN\Xi$, $N\Xi\Xi$ and $NN\Xi\Xi$ systems with maximal 
isospin are bound. Since in these states the strong decay 
$N\Xi\to\Lambda\Lambda$ is forbidden by isospin conservation,
these strange few-body systems will be stable under the strong interaction.
These results may suggest that other states with different number of 
$N$'s and $\Xi$'s in the maximal isospin channel could also be bound.
\end{abstract}
\maketitle 

{\it Introduction.$-$} Although $\Lambda$ hypernuclei formed by one or two $\Lambda$'s
bound in nuclei have been studied for a long time~\cite{Gal16,Ric15}
this is not the case for $\Xi$ hypernuclei. Only recently, as a result of
the so-called KISO event~\cite{Naa15}, the binding energy of
a $\Xi^-$ and $^{14}$N was determined to be 4.38 $\pm$ 0.25 MeV.
The lack of experimental data on $\Xi$ hypernuclei can be traced back to the fact
that $\Xi-$nucleus bound states once formed
will immediately decay due to the process 
$N\Xi\to\Lambda\Lambda$, as shown by the KISO event, where
the reaction $\Xi^- + ^{14}{\rm N}\to ^{10}_\Lambda {\rm Be} + ^5_\Lambda {\rm He}$
was observed. In this paper, we address the study of bound states of nucleons 
and $\Xi$'s in the maximal isospin channel, i.e., systems consisting only of 
neutrons and negative $\Xi$'s or protons and neutral $\Xi$'s.
The uniqueness of these systems is a consequence
of the two-body interactions between $NN$, $N\Xi$ 
and $\Xi\Xi$ pairs being all in the isospin 1 channel. Thus,
the strong decay $N\Xi\to\Lambda\Lambda$ is forbidden. 
Therefore, such states, if bound, would be stable under the 
strong interaction.

{\it Two-body interactions.$-$} For the identical pairs,
$NN$ and $\Xi\Xi$, the $S$-wave interaction is in the $^1S_0$
channel due to the Pauli principle, while for the $N\Xi$ pair
both the $^1S_0$ and $^3S_1$ channels contribute. As is
well-known, the $NN$ $^1S_0$ channel is almost bound, with the virtual state 
lying slightly below the $NN$ threshold in the
unphysical sheet. Regarding the two-body
interactions containing  a single $\Xi$, a recent update of the 
Nijmegen ESC08c potential giving account of the pivotal
results of strangeness $-2$ physics, the KISO~\cite{Naa15} and 
the NAGARA~\cite{Tak01} events, was recently released.
The $N\Xi$ $^3S_1$ interaction has a bound state, the so-called
$D^*$ with a binding energy of 1.67 MeV while the $N\Xi$ $^1S_0$ 
interaction is mainly repulsive~\cite{Nag15,Rij16}.
Finally, with respect to the strangeness $-4$ sector, the most recent 
update~\cite{Rij13} shows a $\Xi\Xi$ $^1S_0$ attractive interaction,
although unbound. Note that in earlier versions of the Nijmegen
ESC08c potential~\cite{Sto99} this channel had a bound state. 

Observations like those reported in Ref.~\cite{Naa15}
are interesting. However, in this case microscopic calculations are
impossible and, consequently, their interpretation will be always
afflicted by large uncertainties. Meanwhile the scarce experimental 
information gives rise to ample room for speculation.
The present theoretical investigation of the possible existence 
of few-baryon bound states based on potential models simulating the
experimental data are basic tools to advance in the knowledge of 
the details of the $\Xi N$ and $\Xi\Xi$ interactions.
First, it could help to raise the awareness of the experimentalist 
that it is worthwhile to investigate few-baryon systems, specifically 
because for some quantum numbers such states could be stable. Second, 
it makes clear that strong and attractive $\Xi N$ interactions, like 
those suggested by the ESC08c Nijmegen model,
have consequences for the few-body sector and can be easily tested
against future data. 

Recent preliminary results from lattice QCD suggest an overall 
attractive $\Xi N$ interaction~\cite{Sas15}.
Besides the recent update of the ESC08c Nijmegen model~\cite{Nag15,Rij16}, there are other models
predicting bound states in the $\Xi N$ system previously to the KISO event, such as
the chiral constituent quark model of Ref.~\cite{Car12}.
The possible existence of stable few-body states 
containing a $\Xi N$ two-body subsystem is also suggested by the attractive 
character of the $\Xi\Xi$ interactions for some partial 
waves~\cite{Bea12,Sto99,Mil06,Hai10,Hai15,Nae15,Rij13}. 
Recent results of the HAL QCD Collaboration
about the $\Xi\Xi$ interaction~\cite{Sas14}
suggest that the interaction in the $^1S_0$ partial wave is
presumably not as strong as suggested by the Nijmegen potential.
There are also preliminary studies of the
$\Xi\Xi N$ system~\cite{Bea09} indicating that lattice QCD
calculations of multibaryon systems are now within sight.
However, one should keep in mind that there are other models 
for the $\Xi N$ interaction, like the hybrid 
quark--model based analysis of Ref.~\cite{Fuj07}, the effective field 
theory approach of Ref.~\cite{Hai16}, or even some of the
earlier models of the Nijmegen group~\cite{Sto99} that do not present $\Xi N$
bound states and, in general, the interactions are weakly attractive or repulsive.
Thus, one does not expect that these models will give rise to 
bound states containing a $\Xi N$ subsystem. On the other hand, 
current $\Xi$ hypernuclei studies~\cite{Yam10,Hiy08,Yam01} 
have been performed by means of $\Xi N$ interactions derived from the 
Nijmegen models and thus our study complements such previous works 
for the simplest systems that could be studied.

Following Malfliet and Tjon~\cite{Mal69} we take all the 
two-body interactions to consist of local central Yukawa-type
potentials containing an attractive and a repulsive term, i.e.,
\begin{equation}
V(r)=-A\frac{e^{-\mu_Ar}}{r}+B\frac{e^{-\mu_Br}}{r} \, .
\label{eq21} 
\end{equation}
In the case of the $NN$ $^1S_0$ channel we use
the Malfliet-Tjon model with the parameters given in Ref.~\cite{Gib90}.
If it is assumed that only singlet and triplet $S$-wave contribute in the
two-particle channel, the parametrization used in this work, then set III for the
triplet partial wave and set I for the singlet partial wave, gives a triton 
binding energy of 8.3 MeV~\cite{Mal69}. The effect of the repulsive core on
the singlet two-body channel is crucial to get this results, while the 
repulsion on the triplet two-body channel has almost no effect on the
binding. In fact, if the repulsive core in the singlet partial wave is not considered
the triton becomes overbound (see Table II of Ref.~\cite{Mal70}). Based on predictions 
for separable potentials, Ref.~\cite{Mal69} suggests
that the inclusion of the tensor force in the triplet interaction increases
the triton binding energy by 0.3 MeV. Indeed, this is the result obtained in
Ref.~\cite{Fuj02}, where as can be seen in Table III of that work, a five-channel 
calculation ($S$ and $D$ partial waves) 
differs from a two channel calculation (only $S$ partial waves) by about 0.3 MeV.
The influence of local tensor forces in Malfliet-Tjon Yukawa type
interactions has also been studied in Ref.~\cite{Maf69}, showing that the inclusion of
tensor forces reduces the binding energy of the three-body problem by 1 to
1.5 MeV, depending on the $D$-wave percentage. Thus, the local Yukawa-type potentials
with tensor interaction would give underbinding in the three-body problem
at difference of separable potentials~\cite{Phi68}.
Let us finally mention that it has been demonstrated that separable potentials
can also reproduce the two-body Malfliet-Tjon $T$-matrix,
agreeing with the three-body binding energy to an accuracy
of 2\%~\cite{Har69}.
The parameters of the $N\Xi $ and $\Xi \Xi$ 
channels were obtained by fitting the low-energy data of each channel as given 
in the most recent update of the strangeness $-2$~\cite{Nag15,Rij16} and strangeness 
$-4$~\cite{Rij13} ESC08c Nijmegen potentials. Because we will not 
consider explicitly the coupling to higher mass channels, $\Lambda\Sigma$ and $\Sigma\Sigma$, we may 
loose some binding. Thus, we do not expect that our parametrization of the two-body 
interactions would overestimate the binding energy of the three- and four-body systems.
The low-energy data and the parameters of these models are given in Table~\ref{t1}. 
\begin{table}[t]
\caption{Low-energy parameters of the most recent updates of the ESC08c Nijmegen interactions
for the  $N\Xi$~\cite{Nag15,Rij16} and $\Xi \Xi$~\cite{Rij13} systems, and the parameters of 
the corresponding local potentials given by Eq.~(\ref{eq21}).} 
\begin{ruledtabular} 
\begin{tabular}{ccccccccccc} 
& System & Channel & $a({\rm fm})$ & $r_0({\rm fm})$ & $A$(MeV fm) & 
$\mu_A({\rm fm}^{-1}$) 
& $B$(MeV fm) & $\mu_B({\rm fm}^{-1})$  & \\ \hline
& {$N \Xi$} 
& $^1S_0$ & $0.579$  & $-2.521$ &  $290$  & $3.05$  & $155$ & $1.60$ & \\
& {$N \Xi$} 
& $^3S_1$ & $4.911$  & $0.527$  &  $568$  & $4.56$  & $425$ & $6.73$ & \\ 
& {$\Xi \Xi$} 
& $^1S_0$ & $-7.25$  & $2.00$  &  $155$  & $1.75$  & $490$ & $5.60$ & \\
\end{tabular}
\end{ruledtabular}
\label{t1} 
\end{table}

{\it Results and discussion.$-$} We have obtained the binding energy of the three-body systems 
$NN\Xi$ and $N\Xi\Xi$ by solving the Faddeev equations with the formalism described 
in Ref.~\cite{Gar16} for the case of three fermions when two of them are identical. 
The binding energy of the $NN\Xi\Xi$ system has been derived by using a
variational method with generalized Gaussians detailed in Refs.~\cite{Suz98,Vij09}

We show in Table~\ref{t2} the binding energies of the lightest four systems: $N\Xi$, $NN\Xi$,
$N\Xi\Xi$ and $NN\Xi\Xi$, with maximal isospin. As one can see from this table, the binding 
energy of the maximal isospin few-body systems tends to increase with the number of particles.
However, the increase is not as pronounced as in strangeless nuclei, due to the effect of 
the repulsive $N\Xi$ $^1S_0$ channel as compared to the attractive $NN$ $^1S_0$ channel.
The results shown in Table~\ref{t2} suggest that other maximal isospin systems 
involving a larger number of nucleons and $\Xi$'s might also be bound.
\begin{table}[t]
\caption{Binding energies, $B$, of the
lightest four few-body systems with maximal isospin, $I$.}
\begin{ruledtabular} 
\begin{tabular}{ccccc} 
& System & $(I)J^P$  &  $B$ (MeV)    & \\ \hline
& $N \Xi$ &  $(1)1^+$ &  1.67    & \\
& $NN \Xi$ & $(\frac{3}{2})\frac{1}{2}^+$ 
& 3.00   & \\
& $N\Xi\Xi$ &  
$(\frac{3}{2})\frac{1}{2}^+$ 
& 4.52    & \\
& $NN\Xi \Xi$ &  $(2)0^+$ 
&  7.4   & \\
\end{tabular}
\end{ruledtabular}
\label{t2} 
\end{table}

\acknowledgments 
This work has been partially funded by COFAA-IPN (M\'exico), 
by the Ministerio de Educaci\'on y Ciencia and EU FEDER under 
Contracts No. FPA2013-47443 and FPA2015-69714-REDT,
by Junta de Castilla y Le\'on under Contract No. SA041U16,
and by Generalitat Valenciana PrometeoII/2014/066. 
A.V. is thankful for financial support from the 
Programa Propio XIII of the University of Salamanca.

\end{document}